\documentclass[a4paper,11pt]{article}
\usepackage{amsmath,amsfonts,amssymb,amsthm,amstext,amscd,wasysym}
\usepackage{latexsym}
\usepackage[hidelinks=true]{hyperref}
\usepackage{graphicx}
\usepackage{caption}
\usepackage{comment}
\usepackage{graphicx}
\usepackage{cite}
\usepackage{color}
\usepackage{setspace}
\usepackage{float}
\usepackage[mathscr]{eucal}
\usepackage{xcolor}
\usepackage{chngcntr}

\newtheorem{theorem}{Proposition}

\definecolor{darkred}{rgb}{0.9,0.05,0.05}
\definecolor{darkblue}{rgb}{0.05,0.05,0.6}
\definecolor{darkgreen}{rgb}{0.05,0.6,0.05}
\definecolor{brightgreen}{rgb}{0.1,0.9,0.1}

\renewcommand*{\eqref}[1]{%
  \begingroup
    \hypersetup{
      linkcolor=darkblue,
      linkbordercolor=darkblue,
    }%
    \textcolor{darkblue}{(\ref{#1})}%
  \endgroup 
}

\hypersetup{linkcolor=red,citecolor=brightgreen,urlcolor=brightgreen,colorlinks=true}

\counterwithout{equation}{section}

\begin{document}

\setlength{\skip\footins}{1cm}

\vspace*{2.8cm}
\begin{flushright}\vspace{-5.5cm}
{\small
IPM/P-2016/009 \\
November 10, 2016}\end{flushright}
\vspace{0cm}

\begin{center}

\parbox{1.07\linewidth}{\fontsize{16pt}{20pt}{\hspace*{-0.55cm}\bf{Black Hole Entropy from Entropy of Hawking Radiation}\hspace*{0.65cm}}}\\
 \vspace{9mm}

\centerline{\large{Sajad Aghapour\footnote{e-mail: s\_aghapour@physics.sharif.edu}$^{a,b}$ and Kamal Hajian\footnote{e-mail: kamalhajian@ipm.ir}$^a$}}

\vspace{3mm}
\normalsize
 $^a$\textit{School of Physics, Institute for Research in Fundamental
Sciences (IPM), \\P.O. Box 19395-5531, Tehran, Iran\\
$^b$\textit{Department of Physics, Sharif University of Technology,\\
 P.O. Box 11365-8639, Tehran, Iran}}
\vspace{5mm}




\begin{abstract}
\noindent
We provide a simple way for calculating the entropy of a Schwarzschild black hole from the entropy of its Hawking radiation. To this end, we show that if a thermodynamic system loses its energy only through the black body radiation, its loss of entropy is always 3/4 of the entropy of the emitted radiation. This proposition enables us to relate the entropy of an evaporating black hole to the entropy of its Hawking radiation. Explicitly, by calculating the  entropy of the Hawking radiation emitted in the full period of evaporation of the black hole, we find the Bekenstein-Hawking entropy of the initial black hole.    

\end{abstract}
\end{center}
\vspace*{0.4cm}


\section{{Introduction}}

In the early 70's, based on some thermodynamic analyses, it was shown that black holes (BHs) possess nonzero entropy \cite{Bekenstein:1973ft,Bardeen:1973gd}. Soon after, in spite of lacking knowledge about the nature of these probable statistical systems, and by a semi-classical calculation, their temperature was also calculated \cite{Hawking:1976rt}, which is known as Hawking temperature. The method, which has been implicitly used for calculation of this temperature, was based on the following idea: instead of investigating the black hole as a thermodynamic system (for which the underlying statistical substrate is unknown),  we can investigate its black body radiation to find its temperature. This method is somehow similar to finding the surface temperature of the Sun by measuring the temperature of the sunshine as its black body radiation. 

In spite of these seminal achievements, and after more than four decades since the realization of BHs as thermodynamic systems, the question about microscopic origin of the BH entropy has not given a well-established answer. Nonetheless, there have been interesting progresses towards answering this question, which have employed somehow different approaches. The first derivation of BH entropy has been in the context of Euclidean canonical \cite{Gibbons:1976ue} (and later microcanonical \cite{Brown:1992bq}) functional formulation for quantum gravity. Another famous approach presented in the subject's literature is trying to provide the missing microscopic description by utilizing the string theory \cite{Strominger:1996sh}. A different approach is studying BHs in the semi-classical regime, while trying to describe their entropy from the statistics of the quantum fields residing around them \cite{'tHooft:1984re}. Parallel to this line of research, there have been attempts to relate entanglement entropy associated to a subset of the Hilbert space of these quantum fields, to the entropy of the background BHs \cite{Bombelli:1986rw,Callan:1994py,Holzhey:1994we}. Another related approach, which is based on calculation of entanglement entropy for some conformal quantum fields residing on boundaries of the spacetime, has drawn attentions to the holographic description for the BH entropy \cite{Ryu:2006bv,Hubeny:2007xt}. Yet, some recent studies based on building classical (and hopefully quantum) covariant phase space for (extremal) BHs, have been carried out in order to realize the BH entropy \cite{CHSS:2015mza,CHSS:2015bca,Compere:2015knw}. The list continues, \emph{e.g.} by the fuzzball proposal \cite{Lunin:2001jy}, modelling the BH horizon with some fluids \cite{Saravani:2012is}, etc. For a nice overall review see Ref. \cite{Wald:2001gf}, and for reviews focusing on some of the approaches mentioned above see Refs. \cite{Mohaupt:2000mj,Solodukhin:2011gn,Nishioka:2009un,Hajian:2015eha,Mathur:2005zp}.           

In this paper, we calculate the entropy of a Schwarzschild BH, based on the same method which Hawking used to calculate the temperature of this BH. {To be more specific, we will focus on the physics of the emitted radiation, in order to find the entropy of the BH.} For convenience, we use the $\hbar=c=1$. 

\section{{Black hole entropy from the Hawking radiation}}
In order to provide intuitions for the reader, we first consider an unknown thermodynamic system, which our knowledge about it is only through its black body radiation. We want to find the temperature and the entropy of such a system. However, finding the temperature is an easy job by studying the spectrum of its thermal radiation. On the other hand, finding the entropy is composed of two steps: 1) relating the entropy loss of the system and the entropy carried outside by the radiation 2)      
finding the entropy of the black body radiation emitted from the beginning of the process of radiation till the end of it. The latter would be the state at which all possible internal energy that the system can radiate is finished. This can be assumed to be the state with zero entropy. Therefore, by the second step, we find the entropy of the overall radiation; and then by the first step, the initial entropy of the system can be found. 

The first step of the two above can be taken by considering the proposition below, which is a general and interesting one\footnote{To our knowledge, this proposition has not been well-appreciated in the literature.}, enabling us to relate the entropy loss of a black body to the entropy of its black body radiation:
\begin{theorem} If a thermodynamic system at equilibrium loses its internal energy only through its black body radiation into the vacuum, then its loss of entropy is (approximately) $\frac{3}{4}$ of the entropy of the emitted radiation\footnote{In $(1,d)$ dimensional spacetime, the factor would be $d/(d+1)$.}.
\label{proposition}   
\end{theorem}
\begin{proof}
Let us denote the internal energy, temperature, and entropy of our thermodynamic system by $U$, $T$, and $S$ respectively. Consider that the system loses an infinitesimal amount of internal energy $\delta U$ and entropy $\delta S$ through black body radiation. According to the first law of thermodynamics, and by the assumption of the proposition, 
\begin{equation}\label{proof 1}
\delta  U=  T \delta  S \;.
\end{equation}
On the other hand, we consider the black body radiation emitted in the vicinity of the surface of the thermodynamic system, (approximately) as a canonical ensemble of photon\footnote{In the proof, the photon gas can be replace by (arbitrary combination of) other   bosonic massless gases. It is because of cancellation of the spin degrees of freedom from both sides of Eq. \eqref{proof 2}.} gas at thermal equilibrium with the system. The internal energy, temperature and entropy of this photon gas can be denoted by $\delta \tilde U$, $\tilde T$, and $\delta \tilde S$ respectively. By the standard statistical mechanics of photon gases (see \emph{e.g.} exercise 23.6 in Ref. \cite{Blundell:2009zz}), these quantities are related by 
\begin{equation}\label{proof 2}
\delta \tilde S=\frac{4}{3}\frac{\delta \tilde U}{\tilde T}\;.
\end{equation}
 In the equilibrium, $\tilde T=T$. Moreover, by the conservation of energy $\delta U=-\delta \tilde U$. Putting them into Eqs. \eqref{proof 1} and \eqref{proof 2}, then $\delta S=-\frac{3}{4}\delta \tilde S$. Letting a finite amount of radiation to be emitted, the equilibrium temperature gradually varies; but this relation for the entropy variations remains always valid. So, denoting the entropy loss of the system during the radiation process by $\Delta S$ and the entropy of the emitted radiation by $\Delta \tilde S$, we will have, by the extensivity  of the entropies, 
\begin{equation}
\Delta S=-\frac{3}{4}\Delta \tilde S \;.
\vspace*{-0.2cm}
\end{equation} 
\end{proof}
In Proposition \ref{proposition}, the factor $4/3$ is an approximated factor. The approximation can originate from different issues: 1) The black body radiation around the surface of an object has some preferred directions of propagation. Hence, considering it as a perfect canonical ensemble is an approximation. 2) Interactions between these particles are ignored. 3) It has been explicitly assumed that the radiation propagates into the vacuum, while one can consider an ambient matter. 4)  The particles in the radiation are considered to be massless.  Among the four approximations above, the first and second ones would be cancelled out from the analysis together with the factor $4/3$, as will be described later. Hence, they are irrelevant  to the analysis. Nonetheless, details of these approximations can be found in Refs. \cite{Zurek:1982zz,Page:1983ug,Page:2004xp}. The third approximation, \emph{i.e.} propagation into the vacuum, would be an acceptable approximation, because we will study evaporation of a Schwarzschild black hole into the vacuum.  The last approximation, which is the radiation of the massless particles, is justified during the process of evaporation, as long as the BH temperature is lower than the energy scales of the massive particles. For the usual black holes with $M\approx M_{\astrosun}$, the temperature is approximately $T_{_\mathrm{H}}\approx 10^{-7}K$ which is well below the mass scales of the known massive particles. However, this approximation might break at the last moments of the evaporation. We will show that the result of the analysis would be insensitive to this breakdown.   

Another important, but relevant hidden assumption in Proposition \ref{proposition} is the flatness of the background metric; in the derivation of the relation \eqref{proof 2} for canonical photon gases, mode expansions in flat spacetime has been used. Hence, in order to make the proposition applicable for the BH thermodynamic system, we assume that in the process of evaporation of a Schwarzschild BH, \emph{propagation} of Hawking radiation is an adiabatic process. In other words, entropy of the Hawking radiation does not change by its propagation towards the infinity. Notice that this assumption is not in contradiction with the increase in the entropy through the \emph{emission} process at the moment of creation of the radiation. Therefore, we can perform the calculations at spatial infinity, and then equate the resulted entropy with the BH entropy. So, the first step in calculation of the entropy for our BH is taken.

The second step is to calculate the entropy of the Hawking radiation emitted during the whole process of evaporation of the BH. Consider a Schwarzschild BH with the initial mass $M_0$ which evaporates until its mass vanishes. Besides, consider a distant observer who resides at spatial infinity. The Hawking radiation reaches her at the Hawking temperature which is related to the remaining mass $M$ of the black hole by $T_{_\mathrm{H}}=\frac{1}{8\pi G M}$ \cite{Hawking:1976rt} {in which $G$ is the Newton constant}. Therefore, she will receive the Hawking radiation until the end of the evaporation of the BH, while the temperature of the radiation raises proportional to the inverse of the remaining mass of the BH. In order to calculate the entropy of the whole of Hawking radiation, we can first calculate the infinitesimal entropy of radiation within a spherical thin shell concentric with the BH, and passing through the position of the observer. Therefore, by adding up the entropy of such shells of radiation, as time passes from receiving the first radiation till the time of receiving the last radiation by the observer, we can find the whole entropy.   

In order to find the entropy in the thin shell mentioned above, let us study the Hawking radiation within it.   At the time when the Hawking temperature of the radiation within the shell is some $T_{_\mathrm{H}}$, we can use the standard formula for the entropy of a photon gas which is 
\begin{equation}\label{S U relation inf}
\delta \tilde S= \frac{4}{3}\frac{\delta \tilde U}{T_{_\mathrm{H}}}\;,
\end{equation}
in which $\delta \tilde U$ is the energy of the Hawking radiation inside the shell. One needs to integrate over all Hawking radiation passing through the shell in the period of time between receiving the first piece of radiation till the last piece of that. This integration can be done by paying attention to the conservation of energy, $\delta \tilde U=-\delta M$. Therefore, the entropy of the whole of Hawking radiation emitted during the BH evaporation process is
\begin{equation}
\tilde S=-\int_{M_0}^\epsilon \frac{4}{3} \frac{\delta M}{T_{_\mathrm{H}}}=\frac{4}{3}\int_\epsilon^{M_0} (8\pi G M)\, \delta M= \frac{4}{3}(4\pi G M_0^2)+ \mathcal{O}(\epsilon^2)\;. 
\end{equation} 
The $\epsilon$ denotes the energy scale at which the semi-classical approximations break down. For the usual black holes it satisfies $\epsilon \ll M_0 $. Therefore, the contribution to the entropy from the last term in the equation above would be negligible and can be dropped. Moreover, it confirms the irrelevance of the radiation of the massive particles to the final result, which can be emitted at the last moments of the evaporation. Using Proposition \ref{proposition}, entropy of the initial BH, being $3/4$ of the entropy of the radiation, can be found to be
\begin{equation}
S=\frac{3}{4}\tilde S= \frac{4\pi (2GM_0)^2}{4G}=\frac{A}{4G}\;,
\end{equation} 
in which $A=4\pi r_s^2$ (with $r_s=2GM_0$ being the Schwarzschild radius) is the area of the horizon of the initial BH. This result is exactly the Bekenstein-Hawking entropy of the initial BH.

\section{{Remarks and conclusion}}
At the end, the following remarks are necessary to be mentioned.
\begin{itemize}
\item In Ref. \cite{Zurek:1982zz}, W. H. Zurek had provided almost similar calculations to find the entropy of Hawking radiation emitted from an evaporated BH; he \emph{assumes} the entropy of BH to be $S=\frac{A}{4G}$. Then, using the Hawking temperature $T_{_\mathrm{H}}=\frac{1}{8\pi G M}$, and dynamical properties of photon gases, he finds  the entropy of Hawking radiation to be approximately $4/3$ of the entropy of initial BH. In spite of his splendid work, he does not well appreciate that this increasing of the entropy via black body radiation by a factor of $4/3$ is true for any thermodynamic system, and is not a peculiar property of the BHs. This issue has been proved in Proposition \ref{proposition}, and adopted to the BH background via adibaticity of propagations plus performing calculations at asymptotic flat region.  Hence in this context, Zurek's work can be considered as a firm check of the Proposition \ref{proposition}. On the other hand, in the analysis presented here, the entropy of Hawking radiation was first calculated  by considering its temperature at infinity to be $T_{_\mathrm{H}}=\frac{1}{8\pi G M}$, in addition to invoking the statistical mechanics of photon gases. Then, by the crucial usage of Proposition \ref{proposition}, the peculiar Hawking temperature in terms of the energy $M$, and very non-trivial relation between the energy $M$ and the geometric area of the horizon, the entropy of the BH was found to be $S=\frac{A}{4G}$. 
\item As it was advertised before, the factor appearing in the Eq. \eqref{S U relation inf} is the same factor in Proposition \ref{proposition}, with exactly the same approximations about directional motion of interacting particles. Hence, this factor cancels out of the analysis, justifying irrelevance of studying these approximations in details. This cancellation clarifies the difference of our analysis with Zurek's analysis: if one wants to calculate the entropy of the black body radiation from the assumed entropy loss of the object, then this factor and approximations in its derivation would play a crucial role. It is simply because the entropy of the radiation would be eventually the entropy loss of the object multiplied by this factor. In contrast, if one calculates the entropy of the radiation by studying its statistical mechanics, and wants to find the entropy loss of the object using her results, then this factor cancels out. Explaining in another way, the goal of the former work has been elaborating the entropy of the Hawking radiation, which would not result to a clear-cut outcome. In contrast, the aim of the analysis here is the entropy of the BH, which yields the area of it as an accurate result. If we  pay enough attention to this subtle but very important issue, then the difference between the two works would become clear, and the irrelevance of accurate determination of the factor $4/3$ would be justified. We strongly encourage the reader to contemplate this issue, in order to find the logic behind this document. However, the independence of the analysis from the specific choice of this factor is a positive aspect of the analysis; it makes the analysis to be applicable when one goes beyond the approximation of canonical non-interacting photon emission.
\item One might be skeptical about triviality of the analysis, because by invoking the Hawking temperature $T_{_\mathrm{H}}=\frac{1}{8\pi G M}$ and assuming the first law to hold for the black hole, one can integrate the black hole entropy to find the final result. Nonetheless, the proposed derivation of Bekenstein-Hawking entropy in this paper is nontrivial. It is because of nontriviality of physical steps for calculating the entropy of the whole radiation, relating it to the entropy of BH, and finally using the peculiar temperature of Hawking radiation accompanied by geometrical nontrivial relation between energy and surface area of the BH. Moreover, the analysis sheds light on similar approaches such as the brick-wall model \cite{'tHooft:1984re}. Specifically, integrating entropies over spacelike volumes, which leads to divergent result around the horizon, is replaced by an integration over timelike volume at spatial infinity. 
\item The analysis shows that invoking the statistical mechanics of the quantum fields around a BH to describe its microstates, eventually would lead to the analysis presented here.  
\item It can be easily checked that considering the gray body factor for the BHs would not alter our analysis, because Eq. \eqref{S U relation inf} remains unchanged.
\item At the end, we mention that the analysis presented here has been motivated by studying number of photons vs. the entropy of Hawking radiation elaborated in Ref. \cite{Dundar:2015axd}.           
\end{itemize}

{To conclude, we emphasized that one can find the entropy of a black body object solely from studying its radiation, if its temperature is known during the time interval of the complete evaporation. Based on this issue, we promoted a method to derive the Bekenstein-Hawking entropy of a Schwarzschild black hole by analyzing the statistical mechanics of its Hawking radiation at infinity. The main inputs of the calculation, in addition to the Hawking temperature, were adiabatic propagation, conservation of the energy, and the specific constraint between the mass and radius of the black hole.  Not surprisingly, the final result matches the area law of the black hole entropy in General Relativity.}

\paragraph{Acknowledgements:} SA thanks Prof. Golshani for his supports. KH thanks high energy group at University of Barcelona for their hospitality. We would also like to thank members of quantum gravity group at IPM, Roberto Emparan, and Hrvoje Nikolic, for their inspiring and helpful discussions.  Besides, we thank Ms. Pileroudi for her kind helps in the period of developing the paper.  This work has been supported by the \emph{Allameh Tabatabaii} Prize Grant of \emph{National Elites Foundation} of Iran and the \emph{Saramadan grant} of the Iranian vice presidency in science and technology.


\end{document}